\documentclass[12pt,letterpaper]{article}
\pdfoutput=1
\usepackage[colorlinks=false,
   linkcolor=red, 
   citecolor=blue,
    filecolor=red,
    urlcolor=red,
    linktoc=all, %%%
    pdfstartview=FitV,
    bookmarksopen=true]{hyperref}
\usepackage[left=2cm,top=1cm,right=3cm,nohead]{geometry}
\usepackage[utf8]{inputenc} 
\usepackage{enumitem,epsfig,latexsym,amsfonts,amsmath,amsthm,amssymb,amsbsy,multirow,slashed,color,mathrsfs,wasysym,textcomp,subfigure,wrapfig,comment,bbold,array,longtable,multirow}
\usepackage{amsmath}
\usepackage{amssymb}
\usepackage{amsthm}
\usepackage{graphicx}
\usepackage{slashed}
\usepackage{setspace}
\usepackage{mathrsfs}
\usepackage{subfigure}
\usepackage[boxsize=0.5em,aligntableaux=center]{ytableau}
\usepackage[bf]{caption}
\usepackage{diagrams}
\usepackage{tikz}
\usetikzlibrary{positioning}
\usetikzlibrary{intersections} 
\newlength{\PicScale}
\usepackage{breqn}
\usepackage[USenglish]{babel}
\usepackage[toc,page]{appendix}
\usepackage{amsmath}
\usepackage{amssymb}
\usepackage{graphicx}
\usepackage{hhline}
\usepackage{cite}
\usepackage[vcentermath]{youngtab}
\usepackage{geometry}
\usepackage{cite}
\usepackage{datetime,graphicx,textcomp}
\usepackage{bbm} 
\hypersetup{
    pdftitle={},
    pdfauthor={},
    pdfsubject={}
} 
\newcommand{\A}{\mathbf{A}}
\newcommand{\B}{\mathbf{B}}
\newcommand{\C}{\mathbf{C}}
\newcommand{\W}{\mathbf{W}}
\newcommand{\X}{\mathbf{X}}
\newcommand{\Y}{\mathbf{Y}}
\newcommand{\U}{\mathbf{U}}
\newcommand{\V}{\mathbf{V}}
\newcommand{\TT}{\mathbf{T}}

\newcommand{\be}{\begin{equation}}
\newcommand{\ee}{\end{equation}}
\newcommand{\bmat}{\left(\!\!\begin{array}}
\newcommand{\emat}{\end{array}\!\!\right)}

\newcommand{\T}{{\rm T}}

\newcommand{\RR}{{\mathbb{R}}}

\newcommand{\ZZ}{{\mathbb{Z}}}

\newcolumntype{M}[1]{>{\centering\arraybackslash}m{#1}}
\newcolumntype{N}{@{}m{0pt}@{}}
\def\nn{\nonumber}
\numberwithin{equation}{section}

\begin{document}

\thispagestyle{empty}
 \begin{flushright}
LMU-ASC 06/18\\
  MPP-2018-24
 \end{flushright}
\vspace{0.5cm}
\begin{center}
{\LARGE \bf{A note on T-folds and $T^3$ fibrations}}
\vskip1.5cm 
%\\ \vspace{0.3cm} 
Ismail Achmed-Zade$^{1}$, Mark J.~D.~Hamilton$^{2}$, Dieter
    L\"ust$^{1,3}$ and Stefano
Massai$^{4}$\\
\vskip0.8cm
\textit{$^{1}$ Arnold Sommerfeld Center for Theoretical Physics\\
Theresienstra\ss e 37, 80333 M\"unchen, Germany}\\
\vskip0.4cm
\textit{$^{2}$ Institute for Geometry and Topology, University of Stuttgart\\
Pfaffenwaldring 57, 70569 Stuttgart, Germany}\\
\vskip0.4cm
\textit{$^{3}$ Max-Planck-Institut f\"ur Physik\\F\"ohringer Ring 6, 80805
  M\"unchen, Germany}
\vskip0.4cm
\textit{$^{4}$ Enrico Fermi Institute, University of
  Chicago \\ 5640 S Ellis Ave, Chicago, IL 60637, USA}
\vspace{1.5cm}
\end{center}

\begin{abstract}

\noindent
We study stringy modifications of $T^3$--fibered manifolds, where the
fiber undergoes a monodromy in the T--duality group. We determine the
fibration data defining such T--folds from a geometric model, by using a
map between the duality group and the group of large diffeomorphisms of
a four-torus. 
We describe the monodromies induced around duality defects where such
fibrations degenerate and we argue that local solutions receive
corrections from the winding sector, dual to the symmetry--breaking
modes that correct semi--flat metrics. 

\end{abstract}

\newpage

%\tableofcontents

\section{Introduction}

In exploring the space of string compactifications it
is practical to consider a boundary of the moduli space where volume
moduli have become very large, and supergravity is the correct
low-energy theory governing the light modes. However, many
interesting string vacua, that populate the interior of the moduli
space, cannot be analyzed in this way. In particular, this
restriction precludes the study of truly stringy geometries, where the
large symmetry group of string theory is expected to modify the notion
of Riemannian geometry. Examples of such compactifications are
constructed by
modifying the familiar semi-flat SYZ fibrations of Calabi-Yau
manifolds \cite{Strominger:1996it}, allowing the torus fiber to
undergo monodromies in the full U-duality group. 
The resulting spaces are usually
referred to as T-folds \cite{Hellerman:2002ax,Hull:2004in,
  Vegh:2008jn} (when the monodromies are restricted to the T-duality
group) or U-folds \cite{Kumar:1996zx,Liu:1997mb,Martucci:2012jk}.

In order to determine if such spaces are good string backgrounds one
needs to have control on the corrections to the supergravity
approximation and to have a microscopic description of the defects where
the semi-flat approximation breaks down. These are non-geometric defects that
induce a monodromy in the duality group
\cite{deBoer:2012ma,Lust:2015yia}. A way to deal with the first
problem is to use string dualities in order relate the T-duality group with
the group of large diffeomorphisms of a manifold that is part of a
known string compactification, in the spirit of F-theory
\cite{Vafa:1996xn}. This can be done, for example, for
T-folds in the heterotic strings \cite{McOrist:2010jw,Malmendier:2014uka,Garcia-Etxebarria:2016ibz}. The duality
map can then be used to compute the low energy dynamics on the
T-duality defects \cite{Font:2016odl,Font:2017cya}. 

So far, the only known examples
of such non-geometric fibrations are six-dimensional and involve a
stringy modification of $T^2$ fibered K3 surfaces, with the exception
of asymmetric orbifold points in the moduli space of $T^3$ fibered
T-folds \cite{Vegh:2008jn}.

In this note we consider an explicit globally well defined example
of a T-fold that admits a $T^3$ fibration, by realizing a subset of the T-duality
group $O(3,3;\mathbb{Z})$ as the group of large diffeomorphisms of a
$T^4$. We use known families of $T^4$ fibered Calabi-Yau manifolds to
construct a family of such T-folds. In the geometric picture, the
local defects are simply Taub-NUT spaces, and get dualized to
non-geometric defects that are T-dual to NS5 branes. Such T-duality
cannot be extended globally because of topological twists in the
global fibration. We also use the above mentioned map to construct a
geometric description of the non-geometric $T^2$ fibrations of
\cite{Hellerman:2002ax}. In order to get to such a geometric model one
needs to add an extra circle, which is related by duality to
the M-theory circle \cite{Vegh:2008jn}. 
We will also argue that the
local physics on non-geometric defects cannot be fully captured by
such geometric constructions, and involve stringy physics related to
the sector of strings winding cycles in the fiber.

While we will restrict to the case of a two-dimensional base, we have
in mind extensions of these models to the interesting case of $T^3$
fibrations over a three-dimensional base. In appendix \ref{app:T3overR3} we briefly
discuss an attempt in this direction.

\section{Monodromy and duality group}\label{sec:monodromy}

A useful way to construct candidate non-geometric string
compactifications is to use an adiabatic fibration of a CFT on a torus $T^d$
over a base $\mathcal{B}$. Any two theories related by a T-duality
transformation of the fiber in $\mathcal{G}_d= O(d,d;\mathbb{Z})$ are gauge
equivalent (see for example \cite{Giveon:1994fu} for a review on
T-duality), and hence it should be possible to allow for large gauge
transformations in $\mathcal{G}_d$. Generically these involve a
non-trivial action on the fiber volume, and so the total space is a non-geometric
T-fold. 
The notion of a T-fold is not rigorous in general, but we will give a precise construction in
special cases, restricting ourselves to $T^3$ bundles.
Following \cite{Lust:2015yia}, we will define T-folds with base
manifold a circle and then extend this definition to spheres with $n$ punctures.

\subsection{Mapping tori for $\mathcal{G}_3$}

The simplest examples of T-folds $\mathcal{X}$ with $T^3$ fibers can be constructed
by modifying the mapping torus for the mapping class group
$SL(3,\mathbb{Z})$. Let us consider a $T^3$ fibration over the closed
interval $[0,1]$ and making an identification as follows:
\begin{equation}
\mathcal{X} = \frac{T^3 \times [0,1]}{(x,0) \sim (\phi(x),1)} \, .
\end{equation}
We refer to $\phi\in SL(3;\mathbb{Z})$ as the monodromy of the
fibration. It acts on $H_1(T^3;\mathbb{Z})$ in the obvious
way. Depending on the conjugacy class of the monodromy, the total
space $\mathcal{X}$ can acquire the structure of a nil- or a
sol-manifold (see for example \cite{2009arXiv0903.2926B}). We pick a Riemannian metric on the total space with line element
\begin{equation}
ds^2 = d \theta^2 + G_{ab}(\theta) dx^a dx^b \, ,\quad a, b = 1,2,3 \, .
\end{equation}
One readily shows that the (smooth) metric satisfies
\begin{equation}
\phi^T G(0) \phi = G(1) \, ,
\end{equation}
where we further restrict ourselves to monodromies $\phi \in SL(3;\ZZ)
\cap \exp \left(\mathfrak{sl}(3;\RR) \right) $.
One then choses a smooth family of metrics $G(\theta)$ on the $T^3$ fibers as follows:
\begin{equation}
G(\theta) = \exp \left( \theta \log \phi \right) \cdot G(0) \equiv \left[\exp \left( \theta \log \phi \right) \right]^T G(0) \left[ \exp \left( \theta \log \phi \right) \right].
\end{equation}
We define a T-fold by generalizing this construction to monodromies in
the T-duality group $\mathcal{G}_3 = O(3,3;\mathbb{Z})$. In order to
make sense of the definition of $\mathcal{X}$ we  specify a metric $G$
and a two-form $B$-field on the total space by defining them on each
$T_{\theta}^3$ fiber over the interval. i.e. we obtain a family of
metrics and two-forms on the fibers $G(\theta)$, $B(\theta)$, $\theta
\in [0,1]$. We restrict $\phi \in O(3,3;\mathbb{Z}) \cap \exp
\left( \mathfrak{o}(3,3;\RR) \right)$ and we define the T-duality
action in terms of the background matrix $E(\theta)  = G(\theta) +
B(\theta) $:
\begin{equation}\label{Etheta}
E(\theta) = \exp \left( \theta \log \phi \right) \cdot E(0) \equiv \frac{X(\theta) E(0) +
  Y(\theta)}{Z(\theta) E(0) + W(\theta)} \, ,
\end{equation}
where
\begin{equation}
\exp \left( \theta \log \phi \right) =
\begin{pmatrix}
X(\theta) & Y(\theta) \\
Z(\theta) & W(\theta)
\end{pmatrix} \, .
\end{equation}
Note that the image of the exponential map $\exp:
\mathfrak{o}(3,3;\RR) \rightarrow O(3,3;\RR)$ is contained in the
subgroup $SO(3,3;\RR)^{+}$. 
Recall that $SO(3,3;\mathbb{Z})$ is generated by the following type of transformations:
\begin{itemize}[leftmargin=*]
\item Large diffeomorphisms.
These are elements of the form
\begin{equation}
\left(
\begin{array}{cc}
(R^{-1})^T & 0 \\
0 & R
\end{array}
\right), \ \ \ R \in GL(3;\mathbb{Z}).
\end{equation}
These act on $E$ by conjugation.

\item $B$-shifts and $\beta$ transformations.
$B$-shifts are of the form
\begin{equation}
\begin{pmatrix}
\mathbb{E}_3 & \Theta \\
0& \mathbb{E}_3
\end{pmatrix} \, , \quad \Theta^T = -\Theta,
\end{equation}
and are just gauge transformations for the $B$-field, $B_{ij} \mapsto B_{ij} + \Theta_{ij}$. 
$\beta$-transformations on the other hand are transpositions of shifts
\begin{equation} \label{betashift}
\begin{pmatrix}
\mathbb{E}_3 & 0\\
 \omega & \mathbb{E}_3
\end{pmatrix} \, , \quad \omega^T = -\omega,
\end{equation}
and they mix the metric and $B$-field. 
 
\item Factorized dualities.
These are of the form
\begin{equation}
\begin{pmatrix}
\mathbb{E}_3 - E_{ii} & E_{ii} \\
E_{ii} & \mathbb{E}_3 - E_{ii}
\end{pmatrix}
\end{equation}
where $E_{ii}$ is an elementary matrix, i.e. it has entries
$(E_{ii})_{kl} = \delta_{ik} \delta_{il}$.
\end{itemize}
Note that for shifts and geometric monodromies one obtains a
well-defined Riemannian manifold $\mathcal{X}$ over $S^1$ with an $H$
flux. We will refer to $\mathcal{X}$ as geometric if the
monodromy $\phi$ is comprised of shifts and diffeomorphisms. Otherwise
we call $\mathcal{X}$ non-geometric.
We will not consider factorized duality as possible monodromies. 
For $T^2$ fibered T-folds, these were recently found to have an
important role in heterotic theory \cite{Garcia-Etxebarria:2016ibz}.

\subsection{Examples} 

We give few simple examples to illustrate the above
construction. Some of the monodromies that we consider will appear
as local models for the global examples we detail in the next section.
Let us consider first the case of $\phi \in SL(3;\ZZ)$. Note that
conjugation of $\phi$ by another element $\psi$ can be compensated for
by a basis transformation of $H_1(T^3;\ZZ)$. This is induced by a
diffeomorphism $\Psi$, with $\Psi_{\ast} = \psi$, so the geometry of
$\mathcal{X}$ is only determined by the conjugacy class of
$\phi$. Unfortunately, unlike the case of $SL(2;\ZZ)$, no explicit
characterization of the conjugacy classes is known for $SL(n;\ZZ), \ n
\geq 3$.
Nonetheless, we can see that elements of a parabolic conjugacy class
give rise to
spaces $\mathcal{X}$ which are nil-manifolds, i.e. quotient of a
nilpotent Lie group by a cocompact lattice.
The simplest example arises from the embedding of three-dimensional
nil-manifolds and their duals. For instance, the following matrices are all conjugate in $SL(3;\ZZ)$:
\begin{equation}\label{parab1}
M_1 =
\begin{pmatrix}
 1 & 1 & 0 \\
 0 & 1 & 0 \\
 0 & 0 & 1 \\
\end{pmatrix} \, ,\quad M_2 =
\begin{pmatrix}
 1 & 1 & 1 \\
 0 & 1 & 0 \\
 0 & 0 & 1 \\
\end{pmatrix} \, ,\quad M_3 =
\begin{pmatrix}
 1 & 0 & 1 \\
 0 & 1 & 1 \\
 0 & 0 & 1 \\
\end{pmatrix}.
\end{equation}
The total space $\mathcal{X}$ with
$\phi = M_1$ is equipped with the metric
\begin{equation}\label{eq:simplenil}
ds^2 = d\theta^2 + dx^2 +  dz^2 + \left( dy +  \theta
  dx \right)^2 \,
\end{equation}
where $(x,y,z)$ are coordinates on the $T^3$ fiber. We have that
$\mathcal{X} = S^1 \times \text{M}_3$, where $\text{M}_3$ is
obtained as a compact quotient of the Heisenberg group. The mapping tori for
the other elements have metrics
\begin{align}
\mathcal{X}_{M_2} &: \quad ds^2 = d\theta^2 + dx^2 + (dy+\theta dx)^2 +
(dz + \theta dx)^2 \, ,\\
\mathcal{X}_{M_3} &: \quad ds^2 = d\theta^2 + dy^2 + (dz+ \theta
  dx+\theta dy)^2 \, . \nonumber
\end{align}
An example of a infinite order element in a distinct conjugacy class
is 
\begin{equation}
M_4 = \begin{pmatrix}
 1 & 1 & 0 \\
 0 & 1 & 1 \\
 0 & 0 & 1 \\
\end{pmatrix}\, .
\end{equation}
The total space $\mathcal{X}$ is a $\text{Nil}_4$-manifold, whose Lie algebra is
determined by the following non-trivial commutators $\mathfrak{g} =
\{ [t_{\theta},t_x] = t_y - t_z/2,[t_{\theta}, t_y] = t_z \}$
. The induced metric is
\begin{equation}
ds^2 = d\theta^2 +dx^2 + (dy +\theta dx )^2 + \left[
  \frac12(\theta^2-\theta)dx + \theta dy + dz\right]^2 \, .
\end{equation}
One can similarly analyze finite order elements, as well as
diffeomorphisms which involve an exponential action on some of the torus
cycles.

One can use the above method to construct examples of non-geometric
spaces $\mathcal{X}$. In this case we rather consider $\theta$ as a
coordinate on the unit interval. Gluing the two ends of the resulting
``mapping cylinder'' only makes sense if one uses a large gauge
transformation in the string duality group.  
The simplest example can be  found by using an element of $O(3,3;\mathbb{Z})$ which is a $\beta$-transformation. These are elements of
the T-duality group of the form \eqref{betashift}.
In $d=2$ the only non-trivial element is $\omega = i a \sigma_2 $
and it corresponds to a monodromy for the complexified K\"ahler modulus $\rho = B + i
\text{vol}$ of the $T^2$ sending
$\rho \rightarrow \frac{\rho}{a \rho +1}$. In $d=3$ we can parametrize
the general monodromy as
\begin{equation}
M_{\omega} = \begin{pmatrix}
\mathbb{E}_3 & 0\\
- \omega & \mathbb{E}_3
\end{pmatrix}\, ,\quad 
\omega =  \begin{pmatrix} 0& c  & -b\\
-c  &0  &a\\
b&-a&0\end{pmatrix} \, .
\end{equation}
This induces a line element and a B-field 
\begin{align}
ds^2 &= d\theta^2 +
\frac{dx^2+dy^2+dz^2}{1+(a^2+b^2+c^2)\theta^2}+\frac{(a\,
  dx+b \, dy+c\, dz)^2\,\theta^2}{1+(a^2+b^2+c^2)\,\theta^2} \, ,\label{dsbshift}\\
B & =
\frac{-c \,dx\wedge
  dy + b\, x\wedge dz -a \, dy\wedge dz}{1+(a^2+b^2+c^2)\,\theta^2} \theta
\, . \nn
\end{align}
Although we lack a proper description of this kind of non-geometric
spaces $\mathcal{X}$, in this case we can obtain a geometric
description by realizing the $\phi$ monodromy as an element of
$SL(4;\mathbb{Z})$ exploiting the accidental isomorphism
$SL(4;\mathbb{R}) \cong Spin(3,3;\mathbb{R})$, that we construct
explicitly in appendix \ref{app:sotosl}. Restricting the double cover $\psi:SL(4;\mathbb{R}) \rightarrow SO(3,3;\mathbb{R})^{+}$ to $SL(4;\mathbb{Z})$ we obtain the preimage
of $M_{\omega}$: 
\begin{equation}
\psi^{-1}(M_{\omega}) = 
\left(
\begin{array}{cccc}
1 & 0 & 0 & 0 \\
0 & 1 & 0 & 0 \\
0 & 0 & 1 & 0 \\
a &  b & c & 1 \\
\end{array}
\right) \subset SL(4;\mathbb{Z}).
\end{equation} 
We see that we have a geometric description in terms of a higher
dimensional geometric space $\mathcal{Y}$ which is a mapping torus for the
diffeomorphism $\psi^{-1}(M_{\omega})$. The latter is a parabolic
element of $SL(4,\mathbb{Z})$ and in fact $\mathcal{Y}$ is a
five-dimensional nil-manifold. In the following section we will use
this map to construct families of pairs $(\mathcal{Y}_{m,n},
\mathcal{X}_{m,n})$ of T-folds  $\mathcal{X}$ and their geometrical
counterparts $\mathcal{Y}$.

\section{Abelian fibrations and T-folds}\label{abelianfibrations}

We have seen that by realizing a class of nil- and sol-manifolds as
mapping tori of a toroidal compactifications, we can obtain
non-geometric modifications of such manifolds by allowing the
monodromy of these mapping tori to be in the T-duality group. 
In this section we will use the restriction of the double cover $Spin(3,3;\mathbb{R}) \cong SL(4;\mathbb{R}) \rightarrow SO(3,3;\mathbb{R})^{+}$ to $SL(4;\mathbb{Z})$ in order to describe a larger class of T-folds. These
are determined by monodromy data that is equivalent
to a $T^4$ fibration whose total space is a Calabi-Yau three-fold. As a byproduct of this
construction we will be able to realize global models in type II
string theory that contain the T-fects of \cite{Lust:2015yia}.

\subsection{The manifolds $\mathcal{Y}_{m,n}$}

We will describe a family of Calabi-Yau three-folds
$\mathcal{Y}_{m,n}$ that admit a $T^4$ fibration. These are described
by a collection of $SL(4;\mathbb{Z})$
monodromies that specifies a particular set of degenerations of the
fiber. Such a description has been detailed in \cite{Donagi:2008ht}, where the
manifolds $\mathcal{Y}_{m,n}$ were constructed as the M-theory lift of
type IIA orientifold backgrounds with fluxes. By interpreting the
mapping class group of the $T^4$ fiber as the
T-duality group of a $T^3$ compactification, we will use the family of
manifolds 
$\mathcal{Y}_{m,n}$ to construct a semi-flat approximation of T-folds
$\mathcal{X}_{m,n}$ that are $T^3$ fibrations with T-duality
monodromies. We will discuss the validity of such an adiabatic
argument in later sections.

Let us consider a family of spaces $\mathcal{Y}_{m,n}$ obtained as
$T^4$ fibrations over a punctured sphere: 
\begin{equation}
\begin{diagram}
T^4 & \rTo & \mathcal{Y}_{m,n} \\
&& \dTo \\
&& \mathbb{CP}^1 \setminus \{p_1,...,p_M\},
\end{diagram}
\end{equation}
where $M = 24 - 4mn > 0$. The $T^4$ fibers degenerate to singular
fibers over every point $p_i$, and locally around each $p_i$,
$\mathcal{Y}_{m,n}$ is a Lefschetz pencil with $T^4$ fibers.
The monodromies of each pencil are given explicitly by the following matrices in $SL(4;\mathbb{Z})$:
\begin{align}
\A &= 
\left(
\begin{array}{cccc}
1&1 &0 &0 \\
0& 1 &0 &0 \\
0& 0 &1 &0\\
0& 0& 0 &1
\end{array}
\right) \, ,\qquad 
\B_1 = 
\left(
\begin{array}{cccc}
 2 & 1 & 0 & m \\
 -1 & 0 & 0 & -m \\
 n & n & 1 & m n \\
 0 & 0 & 0 & 1 \\
\end{array}
\right) \, ,  \label{Xmnmonodromies}\\
\B_2 & = 
\left(
\begin{array}{cccc}
 2 & 1 & 0 & 0 \\
 -1 & 0 & 0 & 0 \\
 0 & 0 & 1 & 0 \\
 0 & 0 & 0 & 1 \\
\end{array}
\right) \, ,\qquad 
\B_3 = 
\left(
\begin{array}{cccc}
 2 & 1 & -m & 0 \\
 -1 & 0 & m & 0 \\
 0 & 0 & 1 & 0 \\
 n & n & -m n & 1 \\
\end{array}
\right) \, , \nonumber \\
\B_4 &= 
\left(
\begin{array}{cccc}
 2 & 1 & -m & m \\
 -1 & 0 & m & -m \\
 n & n & 1-m n & m n \\
 n & n & -m n & m n+1 \\
\end{array}
\right) \, ,\qquad 
\C_1 = 
\left(
\begin{array}{cccc}
 0 & 1 & 0 & -m \\
 -1 & 2 & 0 & -m \\
 n & -n & 1 & m n \\
 0 & 0 & 0 & 1 \\
\end{array}
\right) \, , \nonumber \\
\C_2 &= 
\left(
\begin{array}{cccc}
 0 & 1 & 0 & 0 \\
 -1 & 2 & 0 & 0 \\
 0 & 0 & 1 & 0 \\
 0 & 0 & 0 & 1 \\
\end{array}
\right) \, ,\qquad 
\C_3 = 
\left(
\begin{array}{cccc}
 0 & 1 & m & 0 \\
 -1 & 2 & m & 0 \\
 0 & 0 & 1 & 0 \\
 n & -n & -m n & 1 \\
\end{array}
\right) \, , \nonumber \\
\C_4 & = 
\left(
\begin{array}{cccc}
 0 & 1 & m & -m \\
 -1 & 2 & m & -m \\
 n & -n & 1-m n & m n \\
 n & -n & -m n & m n+1 \\
\end{array}
\right) \, . \nonumber
\end{align}
Note that we use the inverse matrices of those given in
\cite{Donagi:2008ht}. These monodromies provide a factorization of the identity:
\begin{equation}\label{yidfactorization}
\A^{16- 4mn} \B_1 \C_1\B_2 \C_2 \B_3 \C_3 \B_4 \C_4= \mathbb{1} \, .
\end{equation}
As pointed out in \cite{Donagi:2008ht} all monodromies are conjugate
in $SL(4;\mathbb{Z})$ to $\A$,
which implies that the singular fiber is homeomorphic to $T^2 \times
\text{I}_1$, where $\text{I}_1$ denotes the fishtail singularity in the Kodaira
classification of degenerations of elliptic fibrations. We list the
explicit change of basis that brings $\B_4$ and $\C_4$ to this form:
\begin{align}
\A  & = S_{C}^{-1} \C_4 S_{C}
\, , \qquad 
S_C = 
\left(
\begin{array}{cccc}
-1 & 1 & m & -m \\
-1 & 0 & 0 & 0 \\
n & 0 & 1 & 0 \\
n & 0 & 0 & 1 \\
\end{array}
\right)\in SL(4;\mathbb{Z}) \, , \\
\A &= S_B^{-1} \B_4 S_B \, , \qquad 
S_B = 
\left(
\begin{array}{cccc}
1 & 1 & m & -m \\
-1 & 0 & 0 & 0 \\
n & 0 & 1 & 0 \\
n & 0 & 0 & 1 \\
\end{array}
\right) \in SL(4;\mathbb{Z}) \, .\nonumber
\end{align}
There is no global change of basis that transforms all
monodromies into $\A$ simultaneously, so that while the local structure of the fibration
is $K3 \times T^2$, this structure is not preserved globally. This
twisting is parametrized by the integers $(m,n)$.  We point out that the real local geometry is that of a $K3 \times T^2$, but in general
the complex structure does not need to respect this factorization.

If $m=n=0$, we have instead the global factorization $\mathcal{Y}_{0,0} = K3 \times
T^2$. In fact, in this case we find $\B_1 = \B_i \equiv \B$,  $\C_1 = \C_i
\equiv \C$, and there are a total of 24 degenerations. The monodromies are just
the embedding in $SL(4;\mathbb{Z})$ of the standard $\A$, $\B$, $\C$
monodromies (see section \ref{sec:SL2})
\begin{equation}
\A^{16} (\B \C)^4 = (\A^4 \B \C)^4 \, .
\end{equation}
Here the  $\A^4\B \C$ cluster represents the components of a $\text{I}_0$ type
Kodaira singularity. A physical interpretation is that type IIA theory
on
$\mathcal{X}_{0,0}$ is dual to the $T^6/\mathbb{Z}_2$ type IIB
orientifold  (see for example \cite{Schulz:2004tt} for a
detailed discussion).

\subsection{The T-folds $\mathcal{X}_{m,n}$}\label{sec:TfoldsXmn}

We now apply the map from $SL(4;\mathbb{Z})$ to $SO(3,3;\mathbb{Z})$,
reviewed in Appendix \ref{app:sotosl}, in order to obtain a collection of monodromies in $SO(3,3;\mathbb{Z})$, which
factorize the identity. This provides a global model for a T-fold
over $\mathbb{CP}^1$, with $T^3$ fibers. The explicit monodromies are:

\allowdisplaybreaks
\begin{align}
\A \mapsto \W & =
\left(
\begin{array}{cccccc}
 1 & 0 & 0 & 0 & 0 & 0 \\
 -1 & 1 & 0 & 0 & 0 & 0 \\
 0 & 0 & 1 & 0 & 0 & 0 \\
% \hline
 0 & 0 & 0 & 1 & 1 & 0 \\
 0 & 0 & 0 & 0 & 1 & 0 \\
 0 & 0 & 0 & 0 & 0 & 1 \\
\end{array}
\right) \, ,   \label{Tfoldmonodromies} \\
\B_1 \mapsto \X_1 & = 
\left(
\begin{array}{cccccc}
0 & 1 & -n & 0 & m n & m \\
-1 & 2 & -n & -m n & 0 & m \\
0 & 0 & 1 & -m & -m & 0 \\
%\hline
0 & 0 & 0 & 2 & 1 & 0 \\
0 & 0 & 0 & -1 & 0 & 0 \\
0 & 0 & 0 & n & n & 1 \\
\end{array}
\right) \, ,\nonumber \\
\B_2 \mapsto \X_2 &=
\left(
\begin{array}{cccccc}
0 & 1 & 0 & 0 & 0 & 0 \\
-1 & 2 & 0 & 0 & 0 & 0 \\
0 & 0 & 1 & 0 & 0 & 0 \\
%\hline
0 & 0 & 0 & 2 & 1 & 0 \\
0 & 0 & 0 & -1 & 0 & 0 \\
0 & 0 & 0 & 0 & 0 & 1 \\
\end{array}
\right) \, ,\nonumber \\
\B_3 \mapsto \X_3 &=
\left(
\begin{array}{cccccc}
0 & 1 & 0 & 0 & 0 & 0 \\
-1 & 2 & 0 & 0 & 0 & 0 \\
m & -m & 1 & 0 & 0& 0 \\
%\hline
0 & m n & n & 2 & 1 & -m \\
-m n & 0 & -n & -1 & 0 & m \\
-n & n & 0 & 0 & 0 & 1 \\
\end{array}
\right) \, ,\nonumber \\
\B_4 \mapsto \X_4 &=
\left(
\begin{array}{cccccc}
-m n & 1 & -n & 0 & m n & m \\
-1 & 2-m n & -n & -m n & 0 & m \\
m & -m & 1 & -m & -m & 0 \\
%\hline
0 & m n & n & m n+2 & 1 & -m \\
-m n & 0 & -n & -1 & m n & m \\
-n & n & 0 & n & n & 1 \\
\end{array}
\right) \, ,\nonumber \\
\C_1 \mapsto \Y_1 & = 
\left(
\begin{array}{cccccc}
2 & 1 & -n & 0 & m n & m \\
-1 & 0 & n & -m n & 0 & -m \\
0 & 0 & 1 & -m & m & 0 \\
%\hline
0 & 0 & 0 & 0 & 1 & 0 \\
0 & 0 & 0 & -1 & 2 & 0 \\
0 & 0 & 0 & n & -n & 1 \\
\end{array}
\right) \, ,\nonumber \\
\C_2 \mapsto \Y_2 & =
\left(
\begin{array}{cccccc}
2 & 1 & 0 & 0 & 0 & 0 \\
-1 & 0 & 0 & 0 & 0 & 0 \\
0 & 0 & 1 & 0 & 0 & 0 \\
%\hline
0 & 0 & 0 & 0 & 1 & 0 \\
0 & 0 & 0 & -1 & 2 & 0 \\
0 & 0 & 0 & 0 & 0 & 1 \\
\end{array}
\right) \, ,\nonumber \\
\C_3 \mapsto \Y_3 & =
\left(
\begin{array}{cccccc}
2 & 1 & 0 & 0 & 0 & 0 \\
-1 & 0 & 0 & 0 & 0 & 0 \\
-m & -m & 1 & 0 & 0 & 0 \\
%\hline
0 & m n & -n & 0 & 1 & m \\
-m n & 0 & -n & -1 & 2 & m \\
n & n & 0 & 0 & 0 & 1 \\
\end{array}
\right) \, ,\nonumber \\
\C_4 \mapsto \Y_4 &=
\left(
\begin{array}{cccccc}
2-m n & 1 & -n & 0 & m n & m \\
-1 & -m n & n & -m n & 0 & -m \\
-m & -m & 1 & -m & m & 0 \\
%\hline
0 & m n & -n & m n & 1 & m \\
-m n & 0 & -n & -1 & m n+2 & m \\
n & n & 0 & n & -n & 1 \\
\end{array}
\right) \, .\nonumber
\end{align}
Clearly, all these monodromies are conjugate to $\W$, as they are
in the image of the conjugacy class of $\A$ under a
homomorphism.
We now give a brief interpretation of the degenerations associated
with these monodromies. We first notice that the identity
\begin{equation}
\W^{16-4mn} \X_1\Y_1 \X_2 \Y_2 \X_3 \Y_3 \X_4 \Y_4 = \mathbb{1},
\end{equation}
is satisfied, and hence the charges of all individual defects cancel
globally.
Secondly, the $SO(3,3;\mathbb{Z})$ monodromies come in pairs
$(\X_i,\Y_i)$, which are subject to the same interpretation. 
Having this list at our disposal it is immediate that the
pair $(X_2,Y_2)$ in \eqref{Tfoldmonodromies} are diffeomorphisms.
A calculation shows that both $\X_1$ and $\Y_1$ are a product of a diffeomorphism and a shift, for instance
\begin{equation}
\X_1 = 
\left(
\begin{array}{cccccc}
 1 & 0 & 0 & 0 & m n & m \\
 0 & 1 & 0 & -m n & 0 & m \\
 0 & 0 & 1 & -m & -m & 0 \\
% \hline
 0 & 0 & 0 & 1 & 0 & 0 \\
 0 & 0 & 0 & 0 & 1 & 0 \\
 0 & 0 & 0 & 0 & 0 & 1 \\
\end{array}
\right)
\left(
\begin{array}{cccccc}
 0 & 1 & -n & 0 & 0 & 0 \\
 -1 & 2 & -n& 0 & 0 & 0 \\
 0 & 0 & 1 & 0 & 0 & 0 \\
 %\hline
 0 & 0 & 0 & 2 & 1 & 0 \\
 0 & 0 & 0 & -1 & 0 & 0 \\
 0 & 0 & 0 & n & n & 1 \\
\end{array}
\right).
\end{equation}
\ \\
Similarly $(\X_3,\Y_3)$ are compositions of a $\beta$-transformation and a diffeomorphism, e.g.
\begin{equation}
\X_3 = 
\left(
\begin{array}{cccccc}
 0 & 1 & 0 & 0 & 0 & 0 \\
 -1 & 2 & 0 & 0 & 0 & 0 \\
 m & -m & 1 & 0 & 0 & 0 \\
% \hline
 0 & 0 & 0 & 2 & 1 & -m \\
 0 & 0 & 0 & -1 & 0 & m \\
 0 & 0 & 0 & 0 & 0 & 1 \\
\end{array}
\right)
\left(
\begin{array}{cccccc}
 1 & 0 & 0 & 0 & 0 & 0 \\
 0 & 1 & 0 & 0 & 0 & 0 \\
 0 & 0 & 1 & 0 & 0 & 0 \\
% \hline
 0 & m n & n & 1 & 0 & 0 \\
 -m n & 0 & -n & 0 & 1 & 0 \\
 -n & n & 0 & 0 & 0 & 1 \\
\end{array}
\right) \, .
\end{equation}
\ \\
The interpretation for $(\X_4,\Y_4)$ is slightly more involved.
From a factorization of the corresponding $SL(4,\mathbb{Z})$
monodromies we can write $\C_4$  as a product of a diffeomorphism, 
a $B$-shift, and $\beta$-transformations, and similarly for $\X_4$:
\begin{equation}
\Y_4 = \T^{-1}  \left(
\begin{array}{cccccc}
 1 & 0 & 0 & 0 & mn & m \\
 0 & 1 & 0 & -mn & 0 & -m \\
 0 & 0 & 1 & -m & m & 0 \\
%  \hline
 0 & 0 & 0 & 1 & 0 & 0 \\
 0 & 0 & 0 & 0 & 1 & 0 \\
 0 & 0 & 0 & 0 & 0 & 1 \\
\end{array}
\right) \left(
\begin{array}{cccccc}
 2 & 1 & -n & 0 & 0 & 0 \\
 -1 & 0 & n & 0 & 0 & 0 \\
 0 & 0 & 1 & 0 & 0 & 0 \\
% \hline
 0 & 0 & 0 & 0 & 1 & 0 \\
 0 & 0 & 0 & -1 & 2 & 0 \\
 0 & 0 & 0 & n & -n & 1 \\
\end{array}
\right) T \, ,
\end{equation}

\begin{equation}
 \X_4 = \tilde T^{-1}\left(
\begin{array}{cccccc}
 0 & 1 & 0 & 0 & 0 & 0 \\
 -1 & 2 & 0 & 0 & 0 & 0 \\
 m & -m & 1 & 0 & 0 & 0 \\
% \hline
 0 & 0 & 0 & 2 & 1 & -m \\
 0 & 0 & 0 & -1 & 0 & m \\
 0 & 0 & 0 & 0 & 0 & 1 \\
\end{array}
\right) \left(
\begin{array}{cccccc}
 1 & 0 & 0 & 0 & 0 & 0 \\
 0 & 1 & 0 & 0 & 0 & 0 \\
 0 & 0 & 1 & 0 & 0 & 0 \\
% \hline
 0 & m n & n & 1 & 0 & 0 \\
 -m n & 0 & -n & 0 & 1 & 0 \\
 -n & n & 0 & 0 & 0 & 1 \\
\end{array}
\right)  \tilde T,
 \end{equation}
where
\begin{equation}
T= 
\left(
\begin{array}{cccccc}
 1 & 0 & 0 & 0 & 0 & 0 \\
 0 & 1 & 0 & 0 & 0 & 0 \\
 0 & 0 & 1 & 0 & 0 & 0 \\
% \hline
 0 & 1 & 0 & 1 & 0 & 0 \\
 -1 & 0 & 0 & 0 & 1 & 0 \\
 0 & 0 & 0 & 0 & 0 & 1 \\
\end{array} \right)\, ,\quad 
\tilde T = 
\left(
\begin{array}{cccccc}
 1 & 0 & 0 & 0 & -1 & 0 \\
 0 & 1 & 0 & 1 & 0 & 0 \\
 0 & 0 & 1 & 0 & 0 & 0 \\
% \hline
 0 & 0 & 0 & 1 & 0 & 0 \\
 0 & 0 & 0 & 0 & 1 & 0 \\
 0 & 0 & 0 & 0 & 0 & 1 \\
\end{array}
\right) \, .
\end{equation}

We thus see that while locally all the monodromies are related to a
geometric transformation via
an $O(3,3,\mathbb{Z})$ rotation, this is not true globally, and some
of the monodromies act as $\beta$-shifts that mix volume and
B-field, as in \eqref{dsbshift}. Hence, the collection
\eqref{Tfoldmonodromies} specifies a global model of a T-fold with
$T^3$ fibers. In the following, we will illustrate in some details the particular case
$m=n=1$.

\subsection{$\mathcal{X}_{1,1}$ and hyperelliptic fibrations}

In this section we study in some detail the space
$\mathcal{Y}_{1,1}$ and the corresponding T-fold
$\mathcal{X}_{1,1}$. The manifold $\mathcal{Y}_{1,1}$ is defined from the
collection of monodromies \eqref{Xmnmonodromies} with $m=n=1$. There are a total of
20 defects. 
As pointed out in \cite{Donagi:2008ht}, this manifold has an equivalent
description in terms of the Jacobian of a genus-two fibration, which
provides a different way of
geometrizing the T-fold $\mathcal{X}_{1,1}$.
 A very similar
construction appears for $T^2$-fibered T-folds
of heterotic theory when a single Wilson line has non-trivial
monodromies on the base. In this situation one geometrizes the
T-duality group $O(2,3,\mathbb{Z})$ as the mapping class group of a
genus-2 surface $\Sigma_2$. The Jacobian of $\Sigma_2$ is then related to
a physical compactification of F-theory through an adiabatic fibration
of heterotic/F-theory duality
\cite{Malmendier:2014uka,Font:2016odl}. One can then use the general
classification of degenerations of genus-2 fibrations
\cite{Namikawa:1973yq} to collide the 20 defects of
$\mathcal{Y}_{1,1}$, obtaining T-duality defects in $\mathcal{X}_{1,1}$ that are not T-dual
to geometric ones, as in \cite{Font:2016odl}.

We now briefly outline this construction. To each Riemann surface $\Sigma_g$ of genus $g$, one can associate its Jacobian, which is defined to be
\begin{equation}
\text{Jac}(\Sigma_g) := \text{Pic}_0 (\Sigma_g),
\end{equation}
i.e. the subgroup of degree zero divisors. This group can be endowed with the topology
of a torus $T^{2g}$ and in particular to each genus two surface
$\Sigma_2$, one can canonically associate a Jacobian $T^4$. \footnote{In fact
one also has to specify a two-form $\omega$ called
\textit{polarization}, which will not be important for us in the
following.}
The procedure to construct $\mathcal{Y}_{1,1}$ is as follows. Start with a fibration
\begin{equation}
\begin{diagram}
\Sigma_2 & \rTo & \mathcal{S} \\
&& \dTo \\
&& \mathbb{CP}^1 \setminus \Delta,
\end{diagram}
\end{equation}
where $\Delta$ is a finite set of points over which the fibers are
singular with one shrinking cycle, i.e. nodal curves. The total space
is still smooth. Now replace each $\Sigma_2$ with its Jacobian. The
construction of the singular Jacobians requires special care, but is
feasible (for a detailed construction for the nodal genus two curve
see \cite{Altman2007}; see also the excellent lecture notes
\cite{CompJac}). Its topology will be $\text{I}_1 \times T^2$. One can
realize $\mathcal{S}$ as a branched cover of $\mathbb{CP}^1 \times
\mathbb{CP}^1$, which entails choosing a section of $f \in
\mathcal{O}(6) \times \mathcal{O}(2)$. Here one of the factors $\mathbb{CP}^1$ is the original base, the other is (branch) covered by $\Sigma_2$ in the usual manner. 
Indeed this manifold $\mathcal{S}$ is one of
the so-called \textit{Horikawa} surfaces (see for example \cite{Gompf}).
In order to calculate the number of singular fibers we now exploit two formulae for the Euler characteristic of the total space. One is an analog of the Riemann-Hurwitz formula for (complex) surfaces
\begin{equation}
\chi(\mathcal{S}) = 2 \chi(\mathbb{CP}^1 \times \mathbb{CP}^1) - \chi(B),
\end{equation}
where $B = \{ f = 0\}$. As $f$ has bi-degree $(6,2)$ we conclude $\chi(B) = 5$. This yields
\begin{equation}
\chi(\mathcal{S}) = 2 \cdot 4 + 8 = 16.
\end{equation}
The other formula can be derived by choosing a suitable subdivision of the fibration (in Euclidean topology):
\begin{equation}\label{TopEuler}
\chi(\mathcal{S}) = \chi(\mathbb{CP}^1) \chi(\Sigma_2) + n_{\text{sing}} \left(\chi(\hat{\Sigma}_2) - \chi(\Sigma_2) \right).
\end{equation}
Here $\hat{\Sigma}_2$ is a singular genus $2$ surface with one shrinking cycle. 
Now \eqref{TopEuler} reduces to
\begin{equation}
16 = \chi(\mathcal{S}) = 2 \cdot (-2) + n_{\text{sing}} (-1 - (-2)) = -4 + n_{\text{sing}}.
\end{equation}
This gives the number of singular fibers of the $\Sigma_2$ fibration as
$n_{\text{sing}} = 20$, in agreement with the number of T-fects of
$\mathcal{Y}_{1,1}$. This also agrees with the analysis of
\cite{Malmendier:2014uka,Font:2016odl}. 
\begin{figure}[t!]
\begin{center}
\vspace{0.3cm}
\includegraphics[scale=0.4]{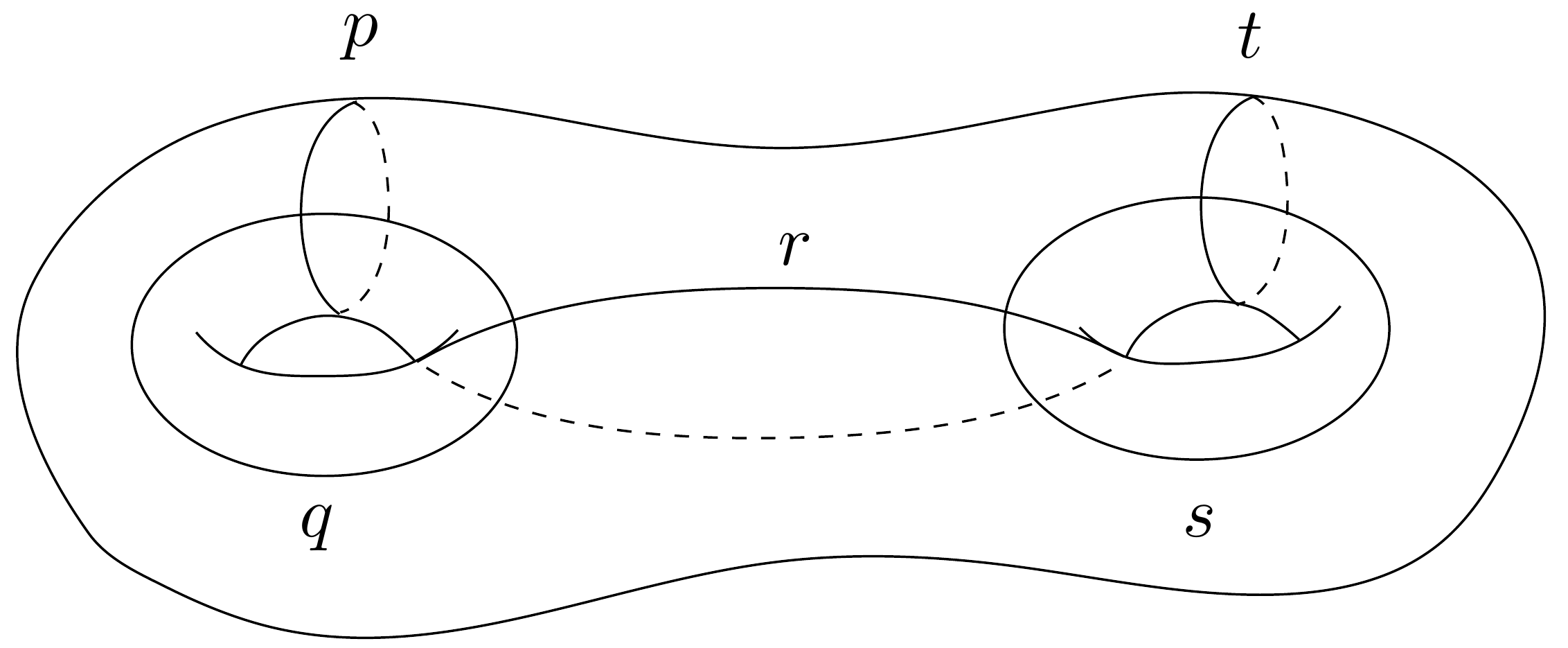}
\caption{The Humphries generators for $\Sigma_2$.}
\label{Fig: Humphries}
\end{center}
\end{figure}
As already mentioned, from the construction of the singular Jacobians one shows
that singular fibers are of type $\text{I}_1 \times T^2$, as we expect from the
fact that all the monodromies that define $\mathcal{Y}_{1,1}$ are
conjugate to the matrix $\A$ in \eqref{Xmnmonodromies}. In fact, one
can see that the list of monodromies  \eqref{Xmnmonodromies} for
$m=n=1$ defines a
set of vanishing cycles for a genus-2 surface by noticing that in that
case, all the matrices are elements of $Sp(4,\mathbb{Z})$, namely
\begin{equation}
\A^t \eta \A = \eta \, ,\quad \B_i^t\eta \B_i =
\eta \, ,\quad \C_i^t \eta \C_i = \eta \, ,
\end{equation}
with 
\begin{equation}
\eta = \begin{pmatrix} 0 & 1 & 0 &0 \\-1 & 0 &0
  &0\\ 0&0&0&1\\ 0&0&-1&0 \end{pmatrix} \, ,
\end{equation}
and they are all conjugate to $\A$ in $Sp(4,\mathbb{Z})$. Note that
$Sp(4,\mathbb{Z}) = \text{Aut}(H_1(\Sigma_2;\mathbb{Z}))$ and from the
surjective map
\begin{equation}
\Phi \, : \, MCG(\Sigma_2)\rightarrow Sp(4,\mathbb{Z}) \, ,
\end{equation}
we see that each monodromy represents an element of the mapping
class group $MCG(\Sigma_2)$, which is in fact a Dehn twist around a
vanishing cycle of $\Sigma_2$. In this case, by a theorem of Humphries
(see for example \cite{mcgprimerbook}), there is a minimum set of
vanishing cycles such that their induced Dehn twists generate all the
mapping class group. For a genus-2 surface these are shown in Figure
\ref{Fig: Humphries}. Picking the basis $(p, q,t,s)$, we see that the corresponding $Sp(4,\mathbb{Z})$ elements are
\begin{align}
\mathbf{P}  = \A &=\begin{pmatrix}
1& 1 &0 &0 \\
0& 1 &0 &0 \\
0& 0 &1 &0\\
0& 0& 0 &1
\end{pmatrix}\, ,\quad \mathbf{Q} =\begin{pmatrix}
1& 0 &0 &0 \\
-1& 1 &0 &0 \\
0& 0 &1 &0\\
0& 0& 0 &1
\end{pmatrix}\, ,\quad \mathbf{R} =\begin{pmatrix}
1& 1 & 0 &-1 \\
0& 1 &0 &0 \\
0& -1 &1 &1\\
0& 0& 0 &1
\end{pmatrix} \, ,\\
\mathbf{S} &= \begin{pmatrix}
1& 0 &0 &0 \\
0& 1 &0 &0 \\
0& 0 &1 &0\\
0& 0& -1 &1
\end{pmatrix} \, , \quad \mathbf{T} = \begin{pmatrix}
1& 0 &0 &0 \\
0& 1 &0 &0 \\
0& 0 &1 &1\\
0& 0& 0 &1
\end{pmatrix} \, .
\end{align}
A global model with trivial monodromy is obtained in this case from
the known relation
\begin{equation}
\mathbf{H}^2 = \mathbb{1} \, ,
\end{equation}
where $\mathbf{H}$ is an hyperelliptic involution, namely a $\pi$ rotation of
$\Sigma_2$ around the horizontal axis in Figure \ref{Fig:
  Humphries}. This is represented by the product
\begin{equation}
\mathbf{H} =
\mathbf{T}\mathbf{S}\mathbf{R}\mathbf{Q}\mathbf{P}\mathbf{P}\mathbf{Q}\mathbf{R}\mathbf{S}\mathbf{T}
\, .
\end{equation}
The relation with the $\A$, $\B_i$, $\C_i$ monodromies arises from the
appropriate braid relations and Hurwitz moves (see for example
\cite{Lust:2015yia} for a review)
\begin{equation}
\B_i = \mathbf{T}_i\mathbf{Q} \mathbf{P} (\mathbf{T}_i \mathbf{Q})^{-1} \, ,\quad \C_i =
\mathbf{T}_i\mathbf{Q}^{-1} \mathbf{P} (\mathbf{T}_i
\mathbf{Q}^{-1})^{-1} \, ,
\end{equation}
where 
\begin{equation}
\mathbf{T}_1  =\begin{pmatrix}
1& 0 &0 &-1 \\
0& 1 &0 &0 \\
0& -1 &1 &0\\
0& 0& 0 &1
\end{pmatrix}\, ,\quad \mathbf{T}_2 = \mathbb{1}\, ,\quad \mathbf{T}_3  =\begin{pmatrix}
1& 0 &1 &0 \\
0& 1 &0 &0 \\
0& 0 &1 &0\\
0& -1& 0 &1
\end{pmatrix}\, ,\quad \mathbf{T}_4  =\begin{pmatrix}
1& 0 &1 &-1 \\
0& 1 &0 &0 \\
0& -1 &1 &0\\
0& -1& 0 &1
\end{pmatrix} \, .
\end{equation}

\section{$SL(2;\mathbb{Z})_{\tau}\times SL(2;\mathbb{Z})_{\rho}$ defects}\label{sec:SL2}

The map between the T-duality group on a $T^3$ and the mapping class
group of a $T^4$ can be used to construct a geometric model for
the class of non-geometric backgrounds introduced in
\cite{Hellerman:2002ax}. Such model is in fact obtained by lifting to
M-theory the U-dual of the semi-flat limit of the latter solutions.
The solutions of \cite{Hellerman:2002ax} are obtained by fibering the
complex and K\"ahler moduli $(\tau, \rho)$ of a two-torus over a
$\mathbb{P}^1$ base. If $\rho$ is fixed one recovers a semi-flat
description of a K3 surface \cite{Greene:1989ya}, while if also $\rho$
varies one obtains a non-geometric modification of the Calabi-Yau
manifold. The metric of the non-trivial space-time directions is
\begin{equation}\label{semiflatmetric}
ds^2 = e^{\varphi} \tau_2 \rho_2dz d\bar z +\frac{\rho_2}{\tau_2} \lvert dx + \tau
dy \rvert^2 
\end{equation}
where $\tau = \tau_1 + i \tau_2$, $\rho = \rho_1 + i \rho_2$ and
$\varphi$ are functions of $z$. At
the generic smooth point in the moduli space, a K3 surface is described
by a torus fibration with 24 singular points of type $\text{I}_1$. 
Locally these degenerations are described by compactified Taub-NUT spaces. In order to obtain a T-fold
$\mathcal{X}$ we need to replace 12 $\text{I}_1$ degenerations with
non-geometric defects determined by a monodromy in $\rho$. This
corresponds to the factorizations
\begin{equation}
\A_{\tau}^8 (\B_{\tau} \C_{\tau})^2 = \mathbb{1} \, ,\quad \A_{\rho}^8 (\B_{\rho} \C_{\rho})^2 = \mathbb{1} 
\end{equation}
where
\begin{equation}\label{eq:ABC}
\A = \begin{pmatrix} 1 & 1\\0& 1 \end{pmatrix} \, , \quad \B
= \begin{pmatrix} 2 & 1\\-1& 0 \end{pmatrix} \, ,\quad \C = \begin{pmatrix} 0 & 1\\-1& 2 \end{pmatrix}
\end{equation}
and the subscript refers to the two factors $SL(2;\mathbb{Z})_{\tau}
\times SL(2;\mathbb{Z})_{\rho}$. It is slightly more useful to use 
two generators of $SL(2;\mathbb{Z})$: $\U = (\A^{-1})^T$, $\V=\A$, that
corresponds to Dehn twists around the $(0,1)$ and $(1,0)$ cycles of
the torus, respectively. The identity then simply factorises
as $(\U\V)^6 = \mathbb{1}$. In order to switch to the
$\A\B\C$ notation one uses the rules: $\U\V\U = \V\U\V$ and $\U\V^n=
\V\TT_n$ with $\TT_{n+2}\TT_n = \B\C$. For example we have
\begin{equation}
(\U\V)^6 = (\V\U\V)^4 = \V^8 \TT_6 \TT_5 \TT_3 \TT_1 = \A^8 (\B\C)^2 \, .
\end{equation}
While the $\A_{\rho}$ monodromy should be associated with a NS5
brane \cite{Ooguri:1995wj}, the $\U$ or $\B$, $\C$ monodromies involve a
non-trivial action on the fiber volume, and this corresponds to a
T-duality defect. The object with monodromy $\U_{\rho}$ is sometimes
referred to as a $5_2^2$ or Q brane
\cite{Obers:1998fb,deBoer:2012ma,Hassler:2013wsa}. 

If we further compactify this setup on a spectator circle, we can
apply the map between $O(3,3;\mathbb{Z})$ and $SL(4;\mathbb{Z})$ to
construct a geometric dual model that involves a geometric $T^4$
fibration, in analogy with the examples discussed in the previous
section. By setting $a = b=0$ in \eqref{eq:sl4column},
\eqref{eq:sl4row} we see that we obtain a global factorization 
\begin{equation}
\begin{diagram}
T^2\times T^2 & \rTo & \mathcal{Y} \\
&& \dTo \\
&& \mathbb{CP}^1 \setminus \{p_1,...,p_{M}\},
\end{diagram}
\end{equation}
where the collections of $\tau$ and $\rho$ monodromies map to the data
that specifies the fibration of the two $T^2$ factors and $M=12+12
=24$.  We see that the four type of elementary degenerations,
corresponding to the type $\text{I}_1$ singularities, NS5 and non-geometric
branes are mapped to the following $SL(4,\mathbb{Z})$ elements:
\begin{equation}
\V_{\tau} \mapsto \begin{pmatrix} \V & 0\\ 0 &\mathbb{1} \end{pmatrix}
\, , \quad \U_{\tau} \mapsto \begin{pmatrix} \U & 0\\ 0 &\mathbb{1} \end{pmatrix}
\, , \quad  \V_{\rho} \mapsto \begin{pmatrix} \mathbb{1} & 0\\ 0 &\V \end{pmatrix}
\, , \quad \U_{\rho} \mapsto \begin{pmatrix} \mathbb{1} & 0\\ 0 &\U\end{pmatrix}
\, . 
\end{equation}
As in the fibrations constructed in section
\ref{abelianfibrations}, locally each degeneration is of type $\text{I}_1\times T^2$, so the
5-branes are lifted to a Taub-NUT space. The global structure is
however different. In the former case for $(m,n)=(0,0)$ one of the $T^2$ factors was trivially
fibered and the total space was simply $\mathcal{Y} = K3 \times
T^2$.

Note that so far we considered T-folds whose geometric description is
a smooth manifold $\mathcal{Y}$. We could consider singular points in the moduli
space obtained by coalescing $\text{I}_1$ degenerations in
$\mathcal{Y}$. This corresponds to coalesce some of the $\tau$ and $\rho$ degenerations. If we only collide $\tau$ or
$\rho$ degenerations separately, the local description of the degeneration will be that of
an ADE singularity in an appropriate duality frame. In particular,
according to the Kodaira table, we can obtain all finite order
elements in $SL(2,\mathbb{Z})$:
\begin{align}
&\text{II}\,  : \, \U \V \, ,\quad \text{III}\, : \, \U \V \U\, ,\quad
              \text{IV}\, : \, (\U \V)^2 \, ,\quad \text{I}_0^{\ast}\,
              : \, (\U \V)^3 \, ,\\
 &\text{IV}^{\ast}\,  : \, (\U \V)^4\, ,\quad \text{III}^{\ast}\, : \, (\U \V)^4\U\, ,\quad \text{II}^{\ast}\, : \, (\U \V)^5 \, ,
\end{align}
as well as the parabolic elements $\text{I}_k\,: \,\,\V^k$,
$\text{I}_k^{\ast} \,:\,\, (\U\V)^3\V^k $.
More interesting examples can be obtained by colliding a $\tau$
and a $\rho$ degeneration, similar to the examples in
\cite{Lust:2015yia,Font:2016odl}. For example, one can consider a
defect of type $[\text{III},\text{III}]$ defined as
\begin{equation}
[\text{III},\text{III}] \, :\,\, \U_{\tau}\V_{\tau}\U_{\tau}
\U_{\rho}\V_{\rho}\U_{\rho} \, .
\end{equation}
In $\mathcal{Y}$, this corresponds to coalesce 6 $\text{I}_1$ mutually
non-local singularities. This is superficially similar to the
heterotic model studied in \cite{Font:2016odl,Font:2017cya}, where a
form of duality was found that, for example, relates a defect of type
$[\text{III},\text{III}]$ with a geometric defect of type
$\text{I}_0^{\ast}$. It would be interesting to see if a similar
result applies to the present models.

\subsection{Quantum corrected metrics}

Both in the example considered in this and the previous sections, all the
local monodromies around the duality defects are conjugate to a simple
Dehn twist around one of the homology cycle of the torus, and in fact
all the degenerations in the geometric spaces $\mathcal{Y}$ are of
type $\text{I}_1 \times T^2$. $\text{I}_1$ is the simplest type of degeneration in
the Kodaira list and corresponds to pinching a cycle of the
torus. This induces a monodromy that is a Dehn twist around the
vanishing cycle. In a geometric space with no flux, a monodromy
factorization such as \eqref{yidfactorization} corresponds to a
list of vanishing cycles for each degenerations. The situation is
different for the spaces $\mathcal{X}$ where the B-field is
non-trivial. The fact that all the monodromies are conjugate to a Dehn
twist just means that we can apply Busher rules in the semi-flat
approximation to exchange the B-field for a non-trivial twist in the
metric. However, it is less clear how to extend such T-duality beyond
the semi-flat approximation. What in the geometric description was a
simple exchange of a vanishing cycle, is now a T-duality in the full
string theory, relating the $\text{I}_1$ singularity with a 5-brane. In order
to describe the local setting, we can neglect the extra circle of the
$T^3$ and just consider a $T^2$ fibration on a disk encircling the
defect. We can take the monodromy of the torus to be, as in \eqref{eq:ABC}
\begin{equation}\label{eq:Vmatrix}
\V = \begin{pmatrix} 1 & 1 \\ 0 & 1  \end{pmatrix} \, ,
\end{equation}
which acts as $\tau \rightarrow \tau +1$ on the complex structure of the torus.
The semi-flat local metric is simply a foliation of the
bundle \eqref{eq:simplenil} and it is given by \eqref{semiflatmetric} with $\rho =
0$ and $\tau = \frac{i}{2\pi}\log (\mu/z)$. The exact metric can be
found by compactifying  a Taub-NUT space on the $(0,1)$ cycle of the
torus, and identifying the shrinking $(1,0)$ cycle with the special
circle. This results in the Ooguri-Vafa metric \cite{Ooguri:1996me}
\begin{equation}\label{eq:OVmetric}
ds^2 = H (dr^2 +r^2d\theta^2 + dx^2)  + \frac{1}{H}(dy + \omega)^2
\end{equation}
with 
\begin{equation}\label{eq:hI0}
H = \frac{1}{2\pi} \log (\mu /r) + \sum_{n\neq 0} e^{i n x}
K_0\left( |n| r\right) \, ,
\end{equation}
where we set the radii to 1 and $K_0$ is the modified Bessel function
of the second kind. The non-perturbative corrections in \eqref{eq:hI0}
localizes the shrinking cycle along the orthogonal one and breaks one
of the $U(1)^2$ isometries of the semi-flat metric. On the other hand,
the action of the monodromy $\V$ on the K\"ahler modulus, i.e. $\rho
\rightarrow \rho + 1$ represents a defect that should be identified
with a NS5 brane \cite{Ooguri:1995wj,Becker:2009df}. The exact metric clearly breaks
both the $U(1)^2$ isometries of the semi-flat solution. In fact after
Poisson resummation the harmonic function can be written as
\begin{equation}\label{eq:NS5loc}
H = \frac{1}{2\pi}\log(\mu/r) + \frac{1}{2\pi} \sum_{k_x, k_y \in
  \mathbb{Z}\setminus \{0\}}K_0(\lambda r) e^{-ik_x x -i k_y y} \, ,
\end{equation}
with $\lambda = \sqrt{k_x^2 + k_y^2}$.
 Hence, by
realizing $\rho$ monodromies as geometric $\text{I}_1$ singularities, we are
missing part of the modes that fully describe the exact metrics beyond the
semi-flat approximation. Similarly, one can consider the non-geometric
monodromies which are $\beta$ transformations in the duality group. For
the $T^2$ example, this is just a monodromy $\U_{\rho}$. Lacking a
worldsheet description of such object we do not know what is the exact
form of the corrected non-geometric solution. One can give the
following argument, which is essentially a semi-flat version of
\cite{Gregory:1997te}. \footnote{See \cite{Harvey:2005ab,Lust:2017jox} for related discussions.} The monodromy $\V_{\rho}$ results in the
non-conservation of momentum along the fiber directions. This is
compensated by an inflow of current where there is a change in the
kinetic terms of the zero modes for translations along the fiber
directions $(x\rightarrow x+ \alpha_x\,  , y\rightarrow
y+\alpha_y)$. Note that $\V_{\rho}$ does not act on the lattice of
windings for strings on the torus. On the other hand, the duality to a
non-geometric monodromy $\U_{\rho}$ results in a trivial action on the
lattice of momenta, but it leads to non-conservation of the winding
numbers $(w_x, w_y)$. The effective dynamics should then involve
 couplings between the winding modes and ``dyonic'' degrees of freedom
 whose kinetic term is increased as the winding charge decreases by
 encircling the defect. This would result in an expression for
 string winding fields that involves Fourier modes similar to
 \eqref{eq:NS5loc}, with the dyonic modes identified with the dual of
 the zero modes $(\alpha_x\, , \alpha_y)$. This structure is not
 visible in supergravity in the non-geometric duality frame, and it is
 presumably accessed by correlation functions in the winding
 sector. We expect this argument to give a qualitatively correct
 picture in a regime where the Bessel function in \eqref{eq:NS5loc} is
 well approximated by exponential decaying terms. Close to the origin,
at least for a stack of defects, one should recover the 5-branes linear dilaton throat.

It is interesting to note that a similar situation arises in the
F-theory models of \cite{McOrist:2010jw,Font:2016odl,Font:2017cya}
that are dual to non-geometric background of the heterotic theory. In that
case, if one describes defects with monodromy in $\tau$ and $\rho$ by
two elliptic fibrations
\begin{equation}
y^2 = x^3 + f_{\tau}(z) x + g_{\tau}(z) \, ,\quad  y^2 =   x^3 +
f_{\rho}(z)  x + g_{\rho}(z) \, ,
\end{equation} 
with $z$ a complex coordinate in the neighborhood of the degeneration,
there exists a map to a dual K3 fibered Calabi-Yau threefold
descending from an adiabatic fibration of 8 dimensional
heterotic/F-theory duality on a common base:
\begin{equation}\label{hetFtheorymap}
y^2 = x^3 - 3f_{\tau}(z)f_{\rho}(z) x u^4 + \frac{\Delta
  _{\tau}(z)\Delta_{\rho}(z)}{16}u^5 - \frac{27}{2} g_{\tau}(z)g_{\rho} (z)
+ u^7 \, ,
\end{equation}
where $\Delta = 4f^3+27 g^2$ is the discriminant of the Weierstra\ss
\:  equations, and $u$ is a complex coordinate on a $\mathbb{P}^1$
base.  Local models of $\text{I}_k$ singularities, NS5 branes and 
non-geometric $\U_{\rho}$ defects are all dualized to the same local
geometric model since the map \eqref{hetFtheorymap} is symmetric in
$\tau$ and $\rho$, as expected from T-duality. The discussion above
implies  a particular form of corrections to the adiabatic
approximation. It would be interesting to check this for NS5 branes,
keeping track of their position on the fiber through the duality.

\vspace{0.5cm}
\subsubsection*{Acknowledgments} 

We are grateful to Valent\'i Vall Camell for useful discussions.
This work was partially supported by the ERC Advanced Grant ``Strings and Gravity''
(Grant.~No. 320045) and by the DFG cluster of excellence ``Origin and
Structure of the  Universe''.   The work of SM is supported in part by DOE grant DE-SC0009924.

\appendix

\section{The map from $SL(4)$ to $SO(3,3)^{+}$}\label{app:sotosl}

We construct the homomorphism of Lie groups \begin{equation}
SL(4;\mathbb{R}) \rightarrow SO(3,3;\mathbb{R})^{+}
\end{equation}
which is a double cover, implying $SL(4;\mathbb{R}) \cong Spin(3,3;\mathbb{R})$. We first pick a basis
\begin{equation}
\mathbb{R}^4 = \langle e_1,...,e_4 \rangle
\end{equation}
which induces a basis of $\Lambda^2 \mathbb{R}^4$ given by
\begin{equation}
\{  e_{23}, - e_{13} , e_{12},e_{14}, e_{24}, e_{34} \}, 
\end{equation}
where $e_{ij} = e_i \wedge e_j$. We define the scalar product on $\Lambda^2 \mathbb{R}^4$ by
\begin{equation}
\langle x, y \rangle e_1 \wedge ... \wedge e_4 = x \wedge y,
\end{equation}
for $x,y \in \Lambda^2 \mathbb{R}^4$.
Now let $A \in SL(4;\mathbb{R})$ act on $\mathbb{R}^4$ by left multiplication. We view elements of $\mathbb{R}^4$ as column vectors. Then there is an induced action of $SL(4;\mathbb{R})$ on $\Lambda^2 \mathbb{R}^4$ given by
\begin{equation}
A \cdot (e_{i} \wedge e_j) = (A e_i) \wedge (A e_j).
\end{equation}
Because of the well-known identity
\begin{equation}
(A e_1) \wedge (A e_2) \wedge (A e_3) \wedge (A e_4) = \text{Det} (A) e_1 \wedge ... \wedge e_4 = e_1 \wedge ... \wedge e_4,
\end{equation}
this action leaves the scalar product on $\Lambda^2 \mathbb{R}^4$ invariant. We therefore expand
\begin{equation}
A \cdot e_{ij} = \sum_{kl} B_{ij, kl} e_{kl},
\end{equation}
and obtain a $6 \times 6$ matrix $B$, which acts on $\Lambda^2 \mathbb{R}^4$ by left multiplication where we view elements of $\Lambda^2 \RR^4$ as column vectors with respect to the basis above. By construction this matrix leaves the scalar product invariant. But explicitly we calculate
\begin{equation}
\langle e_{14}, e_{23} \rangle = 1 \ \ \ \langle e_{24}, -e_{13} \rangle = 1 \ \ \ \langle e_{34}, e_{12} \rangle = 1,
\end{equation}
with all other combinations of basis vectors having vanishing scalar product. In matrix form the scalar product is given by
\begin{equation}
\eta = 
\left(
\begin{array}{ccc|ccc}
0 & 0 & 0 & 1 & 0 & 0 \\
0 & 0 & 0 & 0 & 1 & 0 \\
0 & 0 & 0 & 0 & 0 & 1 \\
\hline 
1 & 0 & 0 & 0 & 0 & 0 \\
0 & 1 & 0 & 0 & 0 & 0 \\
0 & 0 & 1 & 0 & 0 & 0 
\end{array}
\right) \, .
\end{equation}
As mentioned above by construction
\begin{equation}
B^T \eta B = \eta,
\end{equation}
thus $B \in SO(3,3;\mathbb{R})$. 
Now one checks explicitly that
\begin{equation}
\left(
\begin{array}{c|c}
R &  \\
\hline
& 1
\end{array}
\right) \in SL(4;\mathbb{R}) \, ,
\end{equation}
with $R \in SL(3;\mathbb{R})$ is mapped to the diffeomorphism
\begin{equation}
\left(
\begin{array}{c|c}
(R^{-1})^T & 0 \\
\hline
0 & R
\end{array}
\right) \in O(3,3;\mathbb{R}) \, .
\end{equation}
The element
\begin{equation}\label{eq:sl4column}
\left(
\begin{array}{ccc|c}
1& 0 & 0 & a \\
0& 1 & 0 & b \\
0& 0 & 1 & c \\
\hline
0& 0 & 0 & 1 \\
\end{array}
\right)
\end{equation}
maps to
\begin{equation}
\left(
\begin{array}{c|c}
\mathbb{1} & \omega \\
\hline
0 & \mathbb{1}
\end{array}
\right) \, ,
\end{equation}
with
\begin{equation}\label{eq:sl4row}
\omega = \left(
\begin{array}{ccc}
 0 & c & -b \\
-c & 0 & a \\
 b &-a & 0 
\end{array}
\right) \, ,
\end{equation}
which is a gauge transformation for the B-field. Similarly, 
\begin{equation}
\left(
\begin{array}{ccc|c}
1& 0 & 0 & 0 \\
0& 1 & 0 & 0 \\
0& 0 & 1 & 0 \\
\hline
a &b  & c & 1 \\
\end{array}
\right)
\end{equation}
is mapped to a $\beta$-transformation
\begin{equation}
\left(
\begin{array}{c|c}
\mathbb{1} & 0\\
\hline
-\omega & \mathbb{1}
\end{array}
\right) \, .
\end{equation}

\section{SYZ fibrations}\label{app:T3overR3}

\begin{figure}[t!]
\begin{center}
\vspace{0.3cm}
\includegraphics[scale=0.4]{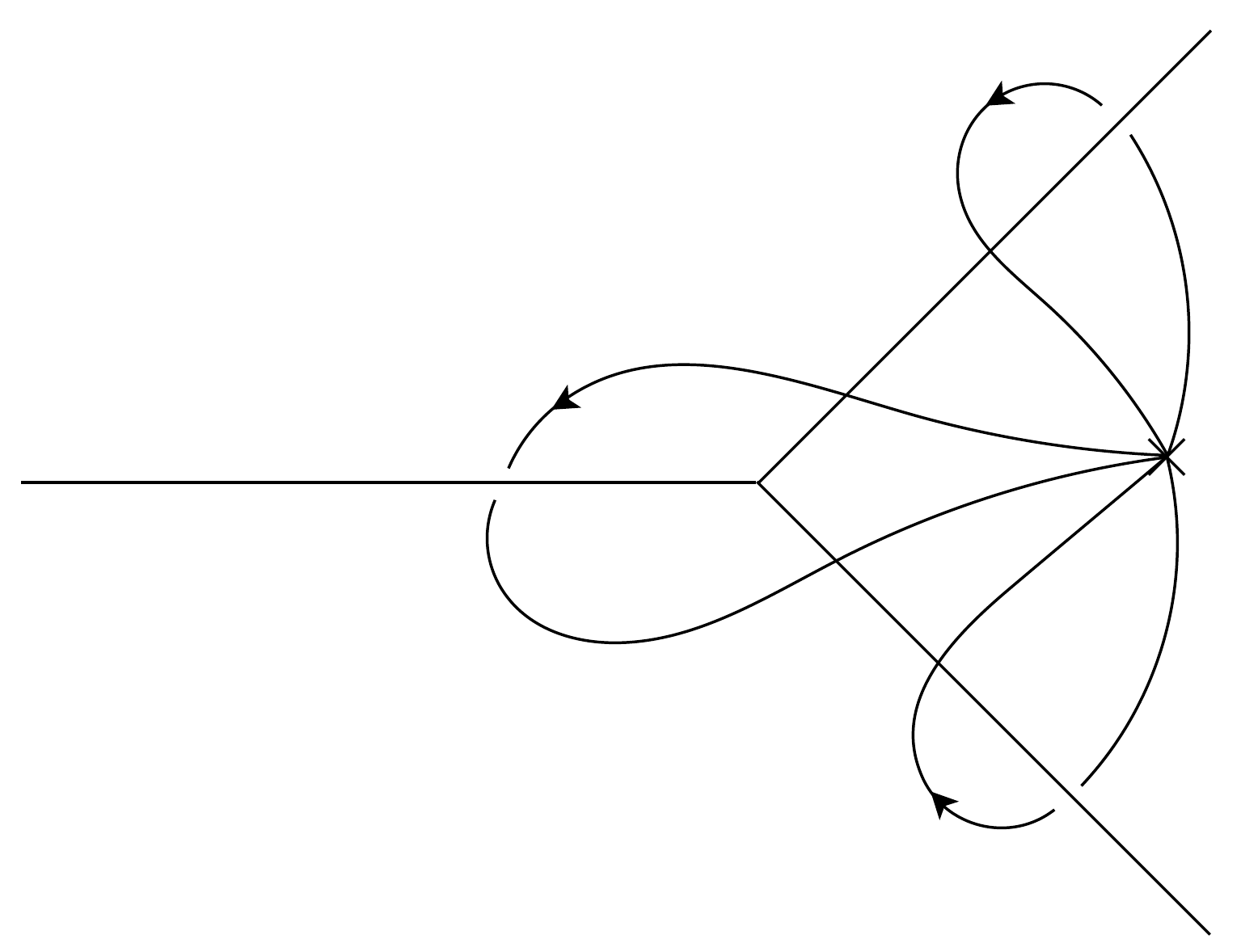}
\caption{Monodromies around a vertex.}
\label{Fig:3loops}
\end{center}
\end{figure}

The extension of our results to the case of a three dimensional base, e.g. $S^3$ are challenging since in this case both local and global aspects are
much less understood, even for the geometric case of SYZ
fibrations. Some non-geometric generalizations corresponding to
asymmetric orbifold points have been considered in
\cite{Vegh:2008jn}. A possibility is that the local structure around
the discriminant locus of a $T^3$ fibrations is modified to account
for non-geometric monodromies. Remember that the quintic viewed as the total space of a $T^3$ fibration has discriminant locus a trivalent graph $\Gamma$ embedded in $S^3$ (see for instance
\cite{2010arXiv1002.4921M} for a review). The monodromy around the
edges of $\Gamma$ is in the same conjugacy class of the matrices in
\eqref{parab1} and the monodromies around a vertex have the following
representatives (see Figure \ref{Fig:3loops}):
\begin{itemize}[leftmargin=*]
\item Positive vertex
\begin{equation}
T_{1+} = \begin{pmatrix} 1&0&1\\0&1&0\\0&0&1 \end{pmatrix}\, , \quad T_{2+}
= \begin{pmatrix} 1&0&0\\0&1&1\\0&0&1\end{pmatrix}\, ,\quad T_{3+} =
T_{2+}^{-1} T_{1+}^{-1} = \begin{pmatrix} 
 1 & 0 & -1 \\
 0 & 1 & -1 \\
 0 & 0 & 1 
\end{pmatrix} \, , 
\end{equation}
\item Negative vertex
\begin{equation}
T_{1-} = \begin{pmatrix} 1&0&1\\0&1&0\\0&0&1 \end{pmatrix}\, , \quad T_{2-}
= \begin{pmatrix} 1&1&0\\0&1&0\\0&0&1\end{pmatrix}\, ,\quad T_{3-} =
T_{2-}^{-1} T_{1-}^{-1} = \begin{pmatrix} 
 1 & -1 & -1 \\
 0 & 1 & 0 \\
 0 & 0 & 1 
\end{pmatrix}  \, .
\end{equation}
\end{itemize}
Since all the monodromies are conjugate to the ones in \eqref{parab1},
it might be possible to extend the conjugacy class in the duality
group, and use the more general monodromies in $O(3,3,\mathbb{Z})$ of
section \eqref{sec:TfoldsXmn}. 
As a first step in this direction, one would like to understand the
analogous of the semi-flat metric \eqref{semiflatmetric} for
$T^3$. We will adapt an approach that was used in \cite{Liu:1997mb}
to study non-perturbative defects with monodromies in the U-duality
group $SL(3,\mathbb{Z})$. Identifying the duality group with the group
of large diffeomorphisms of a $T^3$ this leads to the study of $T^3$
fibered CY three-folds. One start with the following semi-flat ansatz
\begin{equation}\label{metricCY3}
ds^2 = e^{2\phi_1} dx_1^2 +e^{2\phi_2} dx_2^2 +e^{2\phi_3} dx_3^3 +
G_{ij}dy^idy^j \, , \quad G = V^T V
\end{equation}
with $V$ given by 
\begin{equation}
V = e^{-\frac{2\alpha_1 + \alpha_2}{3}}
\left(
\begin{array}{ccc}
1 & a & b \\
0 & e^{-\alpha_1} & e^{-\alpha_1} c\\
0 & 0 & e^{-\alpha_1 - \alpha_2}
\end{array}
\right).
\end{equation}
All the scalars \eqref{metricCY3}
are functions of the $\mathbb{R}^3$ base coordinates $x_i$. We
indicate  by $y_i$ the coordinates on the $T^3$. The prescription of
\cite{Liu:1997mb} is to pick a complex structure by pairing base and
fiber coordinates as follows. We use the differential forms
$dz^i = e^{\phi_i} dx^i + i \delta_{i j}V_{jk} dy^k$, explicitly:
\begin{equation}\begin{aligned}
dz^1 &=e^{\phi_1} dx_1 +i e^{\frac13
  (2\alpha_1+\alpha_2)}(dy_1+a\, dy_2+b\, dy_3) \, ,\\
dz^2 &=e^{\phi_2} dx_2 +i e^{\frac13
  (-\alpha_1+\alpha_2)}(dy_2+c\, dy_3) \, ,\\
dz^3 &=e^{\phi_3} dx_3 +i e^{-\frac13
  (\alpha_1+2\alpha_2)}(dy_3) \, ,
\end{aligned}\end{equation}
and we write
\begin{equation}
J = e^{\phi_i}V_{i
  j}dx^i\wedge dy^j\, ,\quad \Omega = i dz^1\wedge dz^2\wedge dz^3 \, .
\end{equation}
We then see that requiring $d\Omega = dJ = 0$ is equivalent to the
following system of 15 PDEs for the metric moduli:
\begin{equation}\begin{aligned}
\partial_1 a & =  e^{-\alpha_1+\phi_1-\phi_2}
  \partial_2(\alpha_1-\phi_3) \, ,\quad
\partial_2 a  = 2 e^{-\alpha_1-\phi_1+\phi_2}\partial_1 \phi_2\, ,\quad
\partial_3 a = 0 \, ,\\
\partial_1 b & = -2 e^{-\alpha_1-\alpha_2+\phi_1-\phi_3}
\partial_3\phi_1 + c\partial_1 a \, ,\quad
\partial_2 b = c \partial_2 a \, ,\quad
\partial_3 b = 2e^{-\alpha_1-\alpha_2-\phi_1+\phi_3}\partial_1\phi_3
\, ,\\
\partial_1 c & = 0 \, , \quad
\partial_2 c  = -2
e^{-\alpha_2+\phi_2-\phi_3}\partial_3\phi_2 \, , \quad
\partial_3 c  = 2 e^{-\alpha_2-\phi_2+\phi_3}\partial_2\phi_3 \, ,\\
\partial_2\phi_1 & = -\frac13 \partial_2 (2\alpha_1 +\alpha_2) \, ,\qquad
\partial_3 \phi_1 = -\frac13 \partial_3 (2\alpha_1+\alpha_2) \, ,\\
\partial_1\phi_2 & = \frac13\partial_1 (-\alpha_1 +\alpha_2) \, ,\qquad
\partial_3 \phi_2  = \frac13\partial_3 (\alpha_1-\alpha_2) \, ,\\
\partial_1\phi_3 & = \frac13\partial_1 (-\alpha_1 -2\alpha_2) \, ,\qquad
\partial_2\phi_3  = -\frac13\partial_2(\alpha_1+2\alpha_2) \, .
\end{aligned}\end{equation}
By setting for instance $b=c=0$ we can describe the embedding of a
$T^2$ with complex structure $\tau = a + i e^{-\alpha_1}$, and this should be relevant for the monodromy \eqref{parab1}. In this limit
the fields do not depend on $x_3$, and $\phi_3$ is a constant. If we take
$\phi_1=\phi_2$ we then get, fixing an integration constant:
\begin{equation}
\phi_1 = \phi_2 =\alpha_2= -\alpha_1/2 \, ,\quad \partial_1 a = -\partial_2
e^{-\alpha_1}\, ,\quad \partial_2 a = \partial_1 e^{-\alpha_1} \, ,
\end{equation}
the last two equations giving the Cauchy-Riemann equation for $\tau =
a + i e^{-\alpha_1} $ with complex coordinate $z = x_1+i x_2$. The
metric \eqref{metricCY3} takes the form
\begin{equation}
ds^2 = dx_3^2 + dy_3^2+ e^{-\alpha_1} dz d\bar z + G_{ij}dy^idy^j \, , \quad i,j=1,2 \, ,
\end{equation}
with
\begin{equation}
G = e^{\alpha_1}\begin{pmatrix} 1 & a \\a&e^{-2\alpha_1} +
  a^2\end{pmatrix} \, .
\end{equation}
This is the semi-flat metric \eqref{semiflatmetric}, with $\rho = 0$,
where the conformal factor $\varphi$ has been set to zero. This
reproduces the leading order Ooguri-Vafa metric \eqref{eq:OVmetric}
for which
\begin{equation}
\tau =  \frac{i}{2\pi}\log\left(\frac{\mu}{z}\right) \, , \quad
e^{\varphi} = 1 \, .
\end{equation}
The monodromy is $\tau \rightarrow \tau +1$, corresponding to action of
the matrix $\V$ in \eqref{eq:Vmatrix} on $\tau$.
However, we cannot embed a solution for the general conjugacy class of
$\V$, which is parametrized by integers $(p,q)$, since in general this
requires a non-zero $\varphi$. By including the $\rho$ modulus, one
encounter the same situation. The semi-flat approximation of the NS5
brane has $\rho = i/(2\pi)\log(\mu/z)$ and $e^{\varphi} =
1$. The solution for the non-geometric defect with monodromy $\U$ is
given instead by
\begin{equation}
\rho = \frac{2\pi i}{\log\left(\frac{\mu}{z}\right)} \, ,\quad
e^{\varphi} = i \sigma \log\left(\frac{\mu}{z}\right) \, .
\end{equation}
So while we can obtain the correct metric on the fiber, some more work
is needed to write fully non-geometric solutions using this
approach.  We defer a detailed analysis to future work.

%%%%%%%%%%%%%%%%%%%%%%%%%%%%%%%%%%%%%%%%%%%%%%%%%%%%%%%%%%%%%%%%%%%%%%%%%%%

\providecommand{\href}[2]{#2}\begingroup\raggedright\endgroup

%%%%%%%%%%%%%%%%%%%%%%%%%%%%%%%%%%%%
%\end{thebibliography}

\begin{thebibliography}{10}

\bibitem{Strominger:1996it}
A.~Strominger, S.-T. Yau, and E.~Zaslow, ``{Mirror symmetry is T duality},''
  \href{http://dx.doi.org/10.1016/0550-3213(96)00434-8}{{\em Nucl.Phys.}
  {\bfseries B479} (1996) 243--259},
\href{http://arxiv.org/abs/hep-th/9606040}{{\ttfamily arXiv:hep-th/9606040
  [hep-th]}}.
%%CITATION = HEP-TH/9606040;%%.

\bibitem{Hellerman:2002ax}
S.~Hellerman, J.~McGreevy, and B.~Williams, ``{Geometric constructions of
  nongeometric string theories},''
  \href{http://dx.doi.org/10.1088/1126-6708/2004/01/024}{{\em JHEP} {\bfseries
  0401} (2004) 024},
\href{http://arxiv.org/abs/hep-th/0208174}{{\ttfamily arXiv:hep-th/0208174
  [hep-th]}}.
%%CITATION = HEP-TH/0208174;%%.

\bibitem{Hull:2004in}
C.~Hull, ``{A Geometry for non-geometric string backgrounds},''
  \href{http://dx.doi.org/10.1088/1126-6708/2005/10/065}{{\em JHEP} {\bfseries
  0510} (2005) 065},
\href{http://arxiv.org/abs/hep-th/0406102}{{\ttfamily arXiv:hep-th/0406102
  [hep-th]}}.
%%CITATION = HEP-TH/0406102;%%.

\bibitem{Vegh:2008jn}
D.~Vegh and J.~McGreevy, ``{Semi-Flatland},''
  \href{http://dx.doi.org/10.1088/1126-6708/2008/10/068}{{\em JHEP} {\bfseries
  0810} (2008) 068},
\href{http://arxiv.org/abs/0808.1569}{{\ttfamily arXiv:0808.1569 [hep-th]}}.
%%CITATION = ARXIV:0808.1569;%%.

\bibitem{Kumar:1996zx}
A.~Kumar and C.~Vafa, ``{U manifolds},''
  \href{http://dx.doi.org/10.1016/S0370-2693(97)00108-1}{{\em Phys.Lett.}
  {\bfseries B396} (1997) 85--90},
\href{http://arxiv.org/abs/hep-th/9611007}{{\ttfamily arXiv:hep-th/9611007
  [hep-th]}}.
%%CITATION = HEP-TH/9611007;%%.

\bibitem{Liu:1997mb}
J.~T. Liu and R.~Minasian, ``{U-branes and T**3 fibrations},''
  \href{http://dx.doi.org/10.1016/S0550-3213(97)00732-3}{{\em Nucl.Phys.}
  {\bfseries B510} (1998) 538--554},
\href{http://arxiv.org/abs/hep-th/9707125}{{\ttfamily arXiv:hep-th/9707125
  [hep-th]}}.
%%CITATION = HEP-TH/9707125;%%.

\bibitem{Martucci:2012jk}
L.~Martucci, J.~F. Morales, and D.~R. Pacifici, ``{Branes, U-folds and
  hyperelliptic fibrations},''
  \href{http://dx.doi.org/10.1007/JHEP01(2013)145}{{\em JHEP} {\bfseries 1301}
  (2013) 145},
\href{http://arxiv.org/abs/1207.6120}{{\ttfamily arXiv:1207.6120 [hep-th]}}.
%%CITATION = ARXIV:1207.6120;%%.

\bibitem{deBoer:2012ma}
J.~de~Boer and M.~Shigemori, ``{Exotic Branes in String Theory},''
  \href{http://dx.doi.org/10.1016/j.physrep.2013.07.003}{{\em Phys.Rept.}
  {\bfseries 532} (2013) 65--118},
\href{http://arxiv.org/abs/1209.6056}{{\ttfamily arXiv:1209.6056 [hep-th]}}.
%%CITATION = ARXIV:1209.6056;%%.

\bibitem{Lust:2015yia}
D.~L\"ust, S.~Massai, and V.~Vall~Camell, ``{The monodromy of T-folds and
  T-fects},'' \href{http://dx.doi.org/10.1007/JHEP09(2016)127}{{\em JHEP}
  {\bfseries 09} (2016) 127},
\href{http://arxiv.org/abs/1508.01193}{{\ttfamily arXiv:1508.01193 [hep-th]}}.
%%CITATION = ARXIV:1508.01193;%%.

\bibitem{Vafa:1996xn}
C.~Vafa, ``{Evidence for F theory},''
  \href{http://dx.doi.org/10.1016/0550-3213(96)00172-1}{{\em Nucl.Phys.}
  {\bfseries B469} (1996) 403--418},
\href{http://arxiv.org/abs/hep-th/9602022}{{\ttfamily arXiv:hep-th/9602022
  [hep-th]}}.
%%CITATION = HEP-TH/9602022;%%.

\bibitem{McOrist:2010jw}
J.~McOrist, D.~R. Morrison, and S.~Sethi, ``{Geometries, Non-Geometries, and
  Fluxes},'' {\em Adv.Theor.Math.Phys.} {\bfseries 14} (2010) ,
\href{http://arxiv.org/abs/1004.5447}{{\ttfamily arXiv:1004.5447 [hep-th]}}.
%%CITATION = ARXIV:1004.5447;%%.

\bibitem{Malmendier:2014uka}
A.~Malmendier and D.~R. Morrison, ``{K3 surfaces, modular forms, and
  non-geometric heterotic compactifications},''
  \href{http://dx.doi.org/10.1007/s11005-015-0773-y}{{\em Lett. Math. Phys.}
  {\bfseries 105} no.~8, (2015) 1085--1118},
\href{http://arxiv.org/abs/1406.4873}{{\ttfamily arXiv:1406.4873 [hep-th]}}.
%%CITATION = ARXIV:1406.4873;%%.

\bibitem{Garcia-Etxebarria:2016ibz}
I.~Garc{\'\i}a-Etxebarria, D.~L\"ust, S.~Massai, and C.~Mayrhofer, ``{Ubiquity
  of non-geometry in heterotic compactifications},''
  \href{http://dx.doi.org/10.1007/JHEP03(2017)046}{{\em JHEP} {\bfseries 03}
  (2017) 046},
\href{http://arxiv.org/abs/1611.10291}{{\ttfamily arXiv:1611.10291 [hep-th]}}.
%%CITATION = ARXIV:1611.10291;%%.

\bibitem{Font:2016odl}
A.~Font, I.~Garc{\'\i}a-Etxebarria, D.~L\"ust, S.~Massai, and C.~Mayrhofer,
  ``{Heterotic T-fects, 6D SCFTs, and F-Theory},''
  \href{http://dx.doi.org/10.1007/JHEP08(2016)175}{{\em JHEP} {\bfseries 08}
  (2016) 175},
\href{http://arxiv.org/abs/1603.09361}{{\ttfamily arXiv:1603.09361 [hep-th]}}.
%%CITATION = ARXIV:1603.09361;%%.

\bibitem{Font:2017cya}
A.~Font and C.~Mayrhofer, ``{Non-geometric vacua of the
  $\mathbf{\text{Spin}(32)/\mathbb Z_2}$ heterotic string and little string
  theories},'' \href{http://dx.doi.org/10.1007/JHEP11(2017)064}{{\em JHEP}
  {\bfseries 11} (2017) 064},
\href{http://arxiv.org/abs/1708.05428}{{\ttfamily arXiv:1708.05428 [hep-th]}}.
%%CITATION = ARXIV:1708.05428;%%.

\bibitem{Giveon:1994fu}
A.~Giveon, M.~Porrati, and E.~Rabinovici, ``{Target space duality in string
  theory},'' \href{http://dx.doi.org/10.1016/0370-1573(94)90070-1}{{\em
  Phys.Rept.} {\bfseries 244} (1994) 77--202},
\href{http://arxiv.org/abs/hep-th/9401139}{{\ttfamily arXiv:hep-th/9401139
  [hep-th]}}.
%%CITATION = HEP-TH/9401139;%%.

\bibitem{2009arXiv0903.2926B}
C.~{Bock}, ``{On Low-Dimensional Solvmanifolds},'' {\em ArXiv e-prints} (Mar.,
  2009) , \href{http://arxiv.org/abs/0903.2926}{{\ttfamily arXiv:0903.2926
  [math.DG]}}.

\bibitem{Donagi:2008ht}
R.~Donagi, P.~Gao, and M.~B. Schulz, ``{Abelian Fibrations, String Junctions,
  and Flux/Geometry Duality},''
  \href{http://dx.doi.org/10.1088/1126-6708/2009/04/119}{{\em JHEP} {\bfseries
  04} (2009) 119},
\href{http://arxiv.org/abs/0810.5195}{{\ttfamily arXiv:0810.5195 [hep-th]}}.
%%CITATION = ARXIV:0810.5195;%%.

\bibitem{Schulz:2004tt}
M.~B. Schulz, ``{Calabi-Yau duals of torus orientifolds},''
  \href{http://dx.doi.org/10.1088/1126-6708/2006/05/023}{{\em JHEP} {\bfseries
  05} (2006) 023},
\href{http://arxiv.org/abs/hep-th/0412270}{{\ttfamily arXiv:hep-th/0412270
  [hep-th]}}.
%%CITATION = HEP-TH/0412270;%%.

\bibitem{Namikawa:1973yq}
Y.~Namikawa and K.~Ueno, ``The complete classification of fibres in pencils of
  curves of genus two,'' \href{http://dx.doi.org/10.1007/BF01297652}{{\em
  Manuscripta Math.} {\bfseries 9} no.~2, (1973) 143--186}.
  \url{http://dx.doi.org/10.1007/BF01297652}.

\bibitem{Altman2007}
A.~B. Altman and S.~L. Kleiman, {\em The presentation functor and the
  compactified Jacobian},
  \href{http://dx.doi.org/10.1007/978-0-8176-4574-8_2}{pp.~15--32}.
\newblock Birkh{\"a}user Boston, Boston, MA, 2007.
\newblock \url{http://dx.doi.org/10.1007/978-0-8176-4574-8_2}.

\bibitem{CompJac}
J.~Kass, ``Notes on compactified jacobian,'' 2008.
\newblock
  \url{http://people.math.sc.edu/kassj/Lecture%20Notes%20on%20Compactified%20Jacobians.pdf}.

\bibitem{Gompf}
R.~Gompf and A.~Stipsicz, {\em 4-manifolds and Kirby Calculus}.
\newblock Graduate studies in mathematics. American Mathematical Society, 1999.
\newblock \url{https://books.google.de/books?id=ahLKzRUTBbUC}.

\bibitem{mcgprimerbook}
B.~Farb and D.~Margalit, {\em A Primer on Mapping Class Groups}.
\newblock Princeton University Press, 2011.

\bibitem{Greene:1989ya}
B.~R. Greene, A.~D. Shapere, C.~Vafa, and S.-T. Yau, ``{Stringy Cosmic Strings
  and Noncompact Calabi-Yau Manifolds},''
\href{http://dx.doi.org/10.1016/0550-3213(90)90248-C}{{\em Nucl.Phys.}
  {\bfseries B337} (1990) 1}.
%%CITATION = NUPHA,B337,1;%%.

\bibitem{Ooguri:1995wj}
H.~Ooguri and C.~Vafa, ``{Two-dimensional black hole and singularities of CY
  manifolds},'' \href{http://dx.doi.org/10.1016/0550-3213(96)00008-9}{{\em
  Nucl.Phys.} {\bfseries B463} (1996) 55--72},
\href{http://arxiv.org/abs/hep-th/9511164}{{\ttfamily arXiv:hep-th/9511164
  [hep-th]}}.
%%CITATION = HEP-TH/9511164;%%.

\bibitem{Obers:1998fb}
N.~Obers and B.~Pioline, ``{U duality and M theory},''
  \href{http://dx.doi.org/10.1016/S0370-1573(99)00004-6}{{\em Phys.Rept.}
  {\bfseries 318} (1999) 113--225},
\href{http://arxiv.org/abs/hep-th/9809039}{{\ttfamily arXiv:hep-th/9809039
  [hep-th]}}.
%%CITATION = HEP-TH/9809039;%%.

\bibitem{Hassler:2013wsa}
F.~Hassler and D.~L{\"u}st, ``{Non-commutative/non-associative IIA (IIB) Q- and
  R-branes and their intersections},''
  \href{http://dx.doi.org/10.1007/JHEP07(2013)048}{{\em JHEP} {\bfseries 1307}
  (2013) 048},
\href{http://arxiv.org/abs/1303.1413}{{\ttfamily arXiv:1303.1413 [hep-th]}}.
%%CITATION = ARXIV:1303.1413;%%.

\bibitem{Ooguri:1996me}
H.~Ooguri and C.~Vafa, ``{Summing up D instantons},''
  \href{http://dx.doi.org/10.1103/PhysRevLett.77.3296}{{\em Phys.Rev.Lett.}
  {\bfseries 77} (1996) 3296--3298},
\href{http://arxiv.org/abs/hep-th/9608079}{{\ttfamily arXiv:hep-th/9608079
  [hep-th]}}.
%%CITATION = HEP-TH/9608079;%%.

\bibitem{Becker:2009df}
K.~Becker and S.~Sethi, ``{Torsional Heterotic Geometries},''
  \href{http://dx.doi.org/10.1016/j.nuclphysb.2009.05.002}{{\em Nucl.Phys.}
  {\bfseries B820} (2009) 1--31},
\href{http://arxiv.org/abs/0903.3769}{{\ttfamily arXiv:0903.3769 [hep-th]}}.
%%CITATION = ARXIV:0903.3769;%%.

\bibitem{Gregory:1997te}
R.~Gregory, J.~A. Harvey, and G.~W. Moore, ``{Unwinding strings and t duality
  of Kaluza-Klein and h monopoles},'' {\em Adv.Theor.Math.Phys.} {\bfseries 1}
  (1997) 283--297,
\href{http://arxiv.org/abs/hep-th/9708086}{{\ttfamily arXiv:hep-th/9708086
  [hep-th]}}.
%%CITATION = HEP-TH/9708086;%%.

\bibitem{Harvey:2005ab}
J.~A. Harvey and S.~Jensen, ``{Worldsheet instanton corrections to the
  Kaluza-Klein monopole},''
  \href{http://dx.doi.org/10.1088/1126-6708/2005/10/028}{{\em JHEP} {\bfseries
  10} (2005) 028},
\href{http://arxiv.org/abs/hep-th/0507204}{{\ttfamily arXiv:hep-th/0507204
  [hep-th]}}.
%%CITATION = HEP-TH/0507204;%%.

\bibitem{Lust:2017jox}
D.~L{\"u}st, E.~Plauschinn, and V.~Vall~Camell, ``{Unwinding strings in
  semi-flatland},'' \href{http://dx.doi.org/10.1007/JHEP07(2017)027}{{\em JHEP}
  {\bfseries 07} (2017) 027},
\href{http://arxiv.org/abs/1706.00835}{{\ttfamily arXiv:1706.00835 [hep-th]}}.
%%CITATION = ARXIV:1706.00835;%%.

\bibitem{Giveon:1999px}
A.~Giveon and D.~Kutasov, ``{Little string theory in a double scaling limit},''
  \href{http://dx.doi.org/10.1088/1126-6708/1999/10/034}{{\em JHEP} {\bfseries
  10} (1999) 034},
\href{http://arxiv.org/abs/hep-th/9909110}{{\ttfamily arXiv:hep-th/9909110
  [hep-th]}}.
%%CITATION = HEP-TH/9909110;%%.

\bibitem{2010arXiv1002.4921M}
D.~R. {Morrison}, ``{On the structure of supersymmetric T3 fibrations},'' {\em
  ArXiv e-prints} (Feb., 2010) ,
  \href{http://arxiv.org/abs/1002.4921}{{\ttfamily arXiv:1002.4921 [math.AG]}}.

\end{thebibliography}
\end{document}